# On-Chip Scalable Coupled *Single* Photon Emitter- All Dielectric Multifunctional *Quantum* Optical Circuits Working on a Single Collective Mie Resonance

SWARNABHA CHATTARAJ, [1] JIEFEI ZHANG, [2,3] SIYUAN LU,[4] AND ANUPAM MADHUKAR[*,1,2,3]

[1]*Ming Hsieh Department of Electrical Engineering, University of Southern California, Los Angeles, California 90089, USA*
[2]*Department of Physics and Astronomy, University of Southern California, Los Angeles, California 90089, USA*
[3]*Mork Family Department of Chemical Engineering and Materials Science, University of Southern California, Los Angeles, California 90089, USA*
[4]*IBM Thomas J. Watson Research Center, Yorktown Heights, New York, 10598, USA*
*\*madhukar@usc.edu*

**Abstract:** We present finite element method based simulation results for on-chip controlled photon interference in optical networks based upon interconnected arrays of primitives comprising a single photon source (SPS) embedded in lithographically fabricable rectangular dielectric building block (DBB) based metastructures in which a single collective Mie resonance of the *whole circuit* provides the needed light manipulating functions of nanoantenna, state-preserving-guiding, beam-splitting and beam-combining. Efficient coupling of the SPSs to the co-designed collective Mie mode of DBBs is achieved. The findings provide strong incentive for the fabrication and examination of such optical circuits that underlie on-chip scalable nanophotonic quantum information processing systems.

## 1. Introduction

A key requirement for realizing on-chip quantum information processing (QIP) using photonic systems is to establish controllable on-chip photon interference that forms the basis of quantum evolution of multiple photon-qubits and can potentially result in scalable QIP platforms [1,2]. Several proof-of-concept demonstrations of the validity of such platforms have been demonstrated using the well-established silicon photonics technology but the needed single photon source (SPS) is mimicked by off-chip photon-starved laser pulse [3,4], or on-chip parametric down-conversion, or four-wave mixing based classical source [5,6]. The needed scalable integration of on-chip SPSs with light manipulating networks have been elusive to this date. Realization of such on-chip optical circuits with SPSs will thus provide a scalable platform for photon based QIP systems. Equally significant, it will also open path towards manipulating light-matter interaction that forms a dominant component in short and long-distance entanglement in quantum networks [7].

A major factor in the lack of scalability of the currently available SPS-light manipulating unit (LMU) integrated platforms [8, 9, 10] is the lack of spectral uniformity and spatial ordering for the conventionally explored approach of using the lattice-mismatch strain-driven 3D island quantum dots, dubbed self-assembled island quantum dots (SAQDs) [2, 11, 12] as SPSs [12]. Recently efforts have been made to overcome these difficulties by realizing spatially defined quantum structures exploiting surface curvature stress gradient driven adatom migration during epitaxial growth on lithographically patterned III-V semiconductors [12]. Ordered quantum dots in 3-fold symmetric recesses on GaAs(111)B have been developed by the Kapon group [13-16]- an approach that has led to integrable on-chip SPSs for very low lattice mismatched III-V systems. Alternatively, it has been demonstrated that growth on mesas with properly chosen mesa orientation [12] leads to single quantum dot formation on mesa top [17-19] that has been recently demonstrated to be single photon sources with >99% purity [20]. This class of mesa-top single QDs (MTSQDs) allows for defect-free strain accommodation in highly lattice mismatched combinations such as InAs/GaAs as well. It thus allows more versatility in tailoring the emission wavelength and hence can be readily implemented at communication wavelengths ranging from 980nm to 1550nm [18] or other UV, mid IR, etc.) regimes.

As noted before [17, 27], the spatially regular arrays of MTSQD based SPSs open a new path–for the much sought realization of on-chip integrable and scalable quantum optical networks through monolithic integration with on-chip light manipulating elements (LMEs). So far attempts made to integrate an SPS with LMEs are dominantly based on photonic crystal membrane approach [8, 21-26] where the SPS is integrated to only individual cavity or waveguide units. In such an approach, it is critical to ensure spectral and spatial mode matching between the different network components (such as the cavities and waveguides) and the SPSs. In contrast, an integrated SPS- LMU–network based on our recently proposed approach [27-29], depicted in Fig.1, exploits a single collective Mie resonance of an interacting array of dielectric building blocks (DBBs) to provide all the needed light manipulating functions [27-29], i.e. (1) deterministic efficient coupling of the emitter with a resonant cavity to enhance emission rate, (2) efficient directed escape into a waveguide, (3) state-preserving propagation, (4) splitting, and (5) combining photons emitted from different sources to achieve controlled on-chip photon interference without mode matching problems. Unlike other proposed applications of such collective Mie resonances in the literature [30] that focus on individual functions such as propagation in the subdiffraction length scale [31, 32], nanoantenna [33] or even enhancing photoluminescence from random island QDs [34], our approach focuses on the holistic need for realization of a complete optical network and attempts to optimize all the required functions with inherent trade-offs.

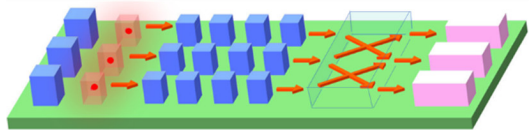

Figure 1. Schematic showing our approach of on-chip DBB array (blue blocks) based multifunctional optical circuit embedded with on-chip SPSs (Red dots) and array of detectors (purple blocks) that exploits on-chip photon interference to create path entanglement.

We note that in Fig. 1, which depicts schematically the conceptual picture of SPS-DBB integrated optical networks based on this holistic approach, not only are the DBBs (blue blocks) taken to be of a generic rectangular shape but so is chosen the shape of the dielectric surrounding the SPSs (red dots inside the red blocks). This mimics one lithographic fabricable realization of this new class of optical networks compatible with existing classes of on-chip SPS-- particularly with the platform of MTSQDs for realizing the SPSs [17]. We also show that exploiting the size of the dielectric around each of the SPS and the location of the SPS within this volume allows us to readily control the coupling of each individual SPS to the collective Mie resonance of the whole circuit—thus enabling control of the afore-mentioned five light manipulating functions for each of the SPSs in the network towards the goal of realizing on-chip interference of photons from distinct on-chip SPSs to create quantum entanglement.

The work presented here is organized as follows: In Section 2 we present a scheme to efficiently couple a single SPS buried in a rectangular DBB with the Mie mode of the same DBB and show how the SPS-single DBB coupling can be readily tailored by adjusting the position of the SPS within the DBB. In Section 3 we extend this to coupling of a single SPS with a single collective Mie resonance of a DBB array that constitutes the nanoantenna and waveguide components. In Section 4 we further extend the approach to include on-chip beam-splitting and beam-combining, thus enabling on-chip interference of photons from distinct SPSs resulting in path entanglement. Section 5 provides some concluding remarks on future directions.

## 2. Coupling Single Photon Source to the Mie Mode of a Single DBB

As expected, realistic simulation of the on-chip integrated structure of an SPS with an LME requires the accounting of the surrounding medium with its shapes and their dielectric effect in affecting the coupling between SPS and LMEs. In this section we thus first address an approach to efficiently and controllably integrate SPS with the DBB based optical circuit. In the SPS-DBB integrated systems as shown in Fig.1, the SPS will invariably be buried in a lithographically carved structure (a mesa). The size and shape of this SPS-bearing block need not be chosen the same as the rest of the DBBs chosen for the co-designed LMU metastructure. In assessing the design of SPS-DBB optical circuits with maximum coupling of the SPS to the DBB-LMU, the lateral extent of the electronic state of the SPS itself is an important consideration in addition to the dielectric effect of the medium of a chosen shape surrounding the SPS. To evaluate this effect, we approximate the transition dipole of the SPS as a point dipole given that the spatial extent of the bound electron and hole states are always <20nm in most contemplated SPSs (quantum dots, defect complexes, etc.), an order of magnitude smaller than the wavelength of interest. In so doing, we can drop the non-locality of the susceptibility [35] of the SPS in our classical Maxwell equations based analysis and use a single point electrical dipole embedded in the DBB to adequately represent the transition dipole moment of the SPS.

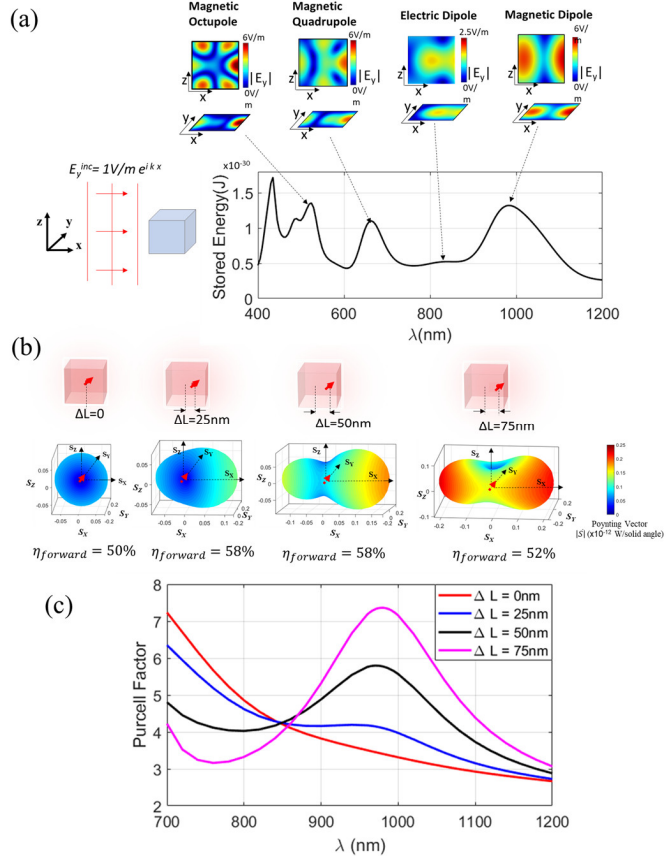

Figure 2. (a) Spectrum of the electromagnetic energy stored in a single DBB of size 220nmx220nmx220nm and refractive index 3.5, surrounded by a uniform refractive index 1.5 upon a plane wave excitation (left inset). The top insets show the E-field distribution of the different Mie resonances. (b) The radiation pattern of an emitter (assumed as a point dipole of strength 1 Debye) embedded in a single DBB for three situations, where the distance of the emitter from the center of the DBB ($\Delta L$) is 0nm, 25nm, 50nm, and 75nm. The directionalities, defined as the percentage of the emitted photons that are in the forward hemisphere (towards +X direction) denoted by $\eta_{forward}$, are respectively 50%, 58%, 58% and 52% for the chosen values of $\Delta L$. (c) The Purcell enhancement as a function of the wavelength of emission for the four cases shown in panel (b).

To optimize the coupling of the SPS transition dipole with the DBB array, we start with analyzing the spectrum of the Mie resonances of a single DBB of the size and shape surrounding the SPS. The rectangular shape of the DBBs is dealt with using finite element method-based simulation with COMSOL Multiphysics. Figure 2(a) shows spectrum of the stored energy within a single DBB of size 220nmx220nmx220nm and refractive index 3.5 representing GaAs, under the condition of a plane wave incidence as shown in the inset (bottom left) of Figure 2(a).

The energy is stored in the supported multiple Mie modes in the single DBB. Hence, the peaks in spectrum of the stored energy in Fig 2(a) represent the spectral locations of the Mie modes. The inset in Fig.2 (a) shows the E-field distribution of these different Mie modes. In particular, the magnetic multipoles are seen as the most prominent peaks. This strong magnetic resonance, a typical characteristic of sub-wavelength scale high index dielectric resonators [30, 36], is exploited, in our case, to establish strong coupling between the modes of the different DBBs as well as with the SPS. Here the magnetic dipole mode is designed to be at 980nm to couple to the transition dipole of the SPS emitting at target wavelength of 980nm [17]. For effectively coupling a Mie mode to a SPS that is buried within the DBB, the position of the SPS transition point dipole should coincide with the E-field maximum of the mode. We use this requirement to maximize the coupling of the SPS transition dipole to the magnetic dipole mode of a single DBB. In Figure 2(b) we show the percentage of emitted photon that are actually emitted in the forward hemisphere (+X direction in this case) (denoted by $\eta_{forward}$) to quantify the directionality imposed by the Mie resonance of the DBB on the SPS emission as a function of the location of the transition point dipole in its own DBB ($\Delta L$, defined with respect to the center of the DBB). We find that the directionality obtains a maximum of 58% when $\Delta L \approx 25nm - 50nm$. In addition to the directionality, we also estimate the effect of the Mie resonance on the emission rate of the SPS, i.e. the Purcell enhancement. The Purcell enhancement is estimated from the E-field distribution by estimating the Green function [27] that leads to:

$$F_p = 1 + \eta\, Im\left(\hat{p}_1 \cdot \frac{\bar{E}_{SPS}(\bar{r}_1)}{|\bar{p}_1|}\right) \ldots\ldots\ldots\ldots\ldots\ldots\ldots\ldots\ldots (1)$$

where $|\bar{p}_1|$ and $\hat{p}_1$ denote, respectively, the magnitude and direction of the electric dipole representing the transition dipole moment of the SPS, $\bar{E}_{SPS}(\bar{r}_1)$ represents the E-field vector generated by the SPS at the location of the SPS itself, and $\eta$ is a constant factor that is given by $\frac{3\,\epsilon_0 c\, \lambda^2}{2\,\pi n_i}$. Here $\lambda$ is the wavelength of light in vacuum, $n_i$ is the refractive index of the DBB, $\epsilon_0$ represents the permittivity of vacuum and c is the free space speed of light. We find that the Purcell enhancement steadily increases as the SPS is brought closer to the surface of the DBB—as shown in Fig.2(c). Guided by the findings in Fig.2(b) and Fig 2(c), we have chosen $\Delta L = 50nm$ for the design of SPS-DBB optical circuits in the next section to maximize the directionality with a reasonable Purcell enhancement of ~5.5 needed to couple the emission from the buried SPS to a collective Mie resonance of the DBB based metastructure designed to manipulate photons from their origination point at the SPS through various positions as they travel through the metastructure (the LMU).

## 3. Coupling a Single Emitter to a Single Collective Mie Mode of a DBB Multifunctional Unit

As known, an array of interacting DBBs with the interaction controlled by the inter-DBB separation can support collective resonances that can be expressed as a linear superposition of the single DBB resonances [27, 29]

$$\Phi_{Collective} = \sum_{i=1}^{N} a_i \phi_i (r - r_i) \ldots\ldots\ldots\ldots\ldots\ldots (2)$$

where $a_i$'s represent the envelop function indicating the spatial distribution of the collective Mie mode over the optical circuit, and $\phi_i(r)$ is the mode of a single DBB. Thus, by enhancing the coupling of the SPS with any of the $\phi_i(r - r_i)$, one can enhance its coupling to the single collective Mie mode of the whole unit designed to provide the desired functionalities. Anticipating the experimental realization demanding protection of the surfaces the entire structure is buried in a uniform medium of refractive index 1.5 mimicking an appropriate protective polymer layer.

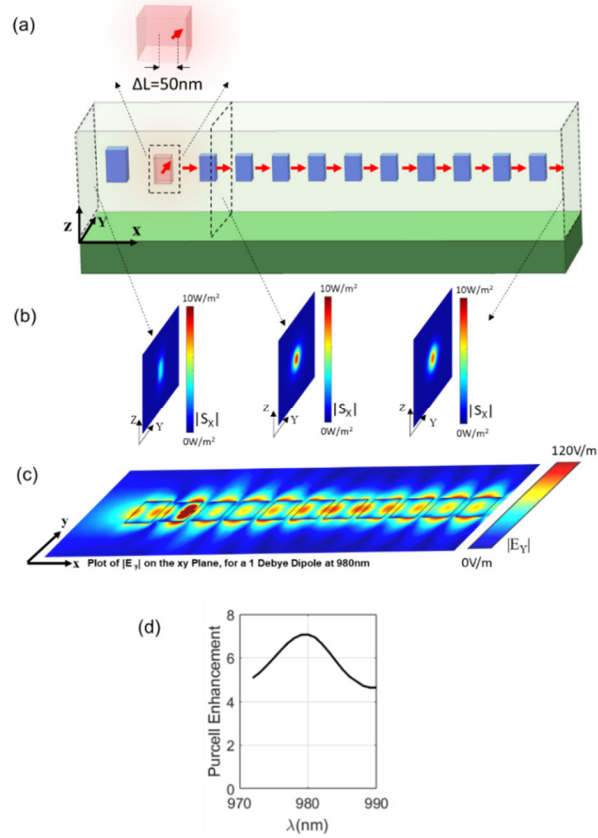

Figure 3. (a) The SPS integrated with the nanoantenna-waveguide unit based on rectangular shaped DBBs. As shown in the inset, the transition dipole representing the SPS is placed at $\Delta L$ =50nm from the center of the DBB. The DBB containing the SPS as well as the DBBs in the waveguide are of size 220nmx220nmx220nm, with a pitch of 275nm. The Reflector DBB of the nanoantenna is of size 220nmx250nmx220nm.(b) The cross-sectional distribution of the Poynting vector calculated for 1Debye transition dipole at three different positions along the axis of the nanoantenna-waveguide. (c) E-field distribution of the collective Mie mode excited by the source dipole emitting at 980nm plotted on the XY plane passing through the center of the DBBs. (d) The Purcell enhancement as a function of wavelength.

The basic design of the nanoantenna-waveguide unit is shown in Figure 3(a). The DBB that embeds the SPS as well as the DBBs of the waveguide part are chosen to be of size 220nmx220nmx220nm, with a pitch of 275nm, and the reflector of the Yagi-Uda nanoantenna (the blue DBB to the left of the SPS) is taken to be of size 220nmx250nmx220nm. This design also ensures that the collective magnetic mode is at 980nm, the targeted emission wavelength of the SPS. The SPS, as discussed in Sec. 2, is embedded at a distance $\Delta L = 50nm$ from the center of the DBB (see inset of Fig. 3(a)). To show the directionality, in panel (b) we plot the distribution of the Poynting vector on cross-sectional planes placed in the forward and backward directions to the SPS as indicated. Approximately 60% of the emitted photons are coupled to the single Mie mode of the waveguide. Notice however that the cross-sectional photon flux remains essentially unchanged over the propagation distance, as depicted by the last panel in Figure 3(b) showing the Poynting vector distribution coming out of the other end of the waveguide. This lossless nature of the photon propagation via the waveguide is reinforced by the plot of the E-field distribution on a XY plane passing through the center of the DBBs as in Fig.3(c) that shows non-diminishing nature of the field distribution over the propagation distance. The same collective mode also enhances the E-field at the location of the SPS—resulting in a Purcell enhancement according to equation (1). For this case the resultant Purcell enhancement spectrum is shown in Figure 3(d). We find that a Purcell enhancement of ~7 is achievable at the target wavelength of 980nm. Thus, the collective Mie resonance of the nanoantenna-waveguide boosted the Purcell enhancement from ~5.5 for the SPS coupled to a single DBB (Fig.2(c)), resulting in shortening of the radiative lifetime of the SPSs thereby improving the photon indistinguishability and enabling photon interference [11,37]. In this illustration, the single collective Mie resonance achieves three needed functions: (1) enhances the emission rate of the SPS, (2) enhances the directionality of the emitted photon and (3) propagates the photon on-chip while maintaining its state, -all in a seamless perfectly mode-matched way. Importantly, such an approach paves the way towards designing scalable integration of multiple SPSs coupled to the same collective Mie mode in larger and more complex network for quantum information processing applications. This is discussed next.

### 4. Splitting and Combining: Photon Interference

The nanoantenna-waveguide metastructure can be readily extended to provide the additional two required functions of on-chip beam-splitting and beam-combining using the same collective Mie resonance of the network [29]. This is illustrated in Fig. 4 which shows the nanoantenna-waveguide-beamsplitting-beamcombining architecture. Such an architecture can be used to guide emission from two distinct SPSs (labelled SPS1 and SPS2 in Fig.4), split the emitted photons on-chip, and recombine them onto a single common branch to create interference between photons originating from two different SPSs—and thus creating path entanglement of the photons. Such an architecture is scalable and can be readily extended to more complex networks that involve a large number N of SPSs at different locations for creating interference and entanglement amongst clusters of SPSs.

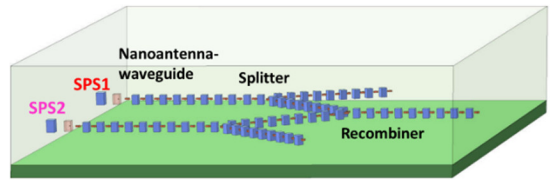

Figure 4. Schematic of the complete structure of nanoantenna-waveguide-splitter-recombiner structure that represent all the five light manipulating functions needed of the optical network, i.e., functions: enhancement of the emission rate of the SPSs, imposing directionality on emission, state-preserving wave-guiding, photon path splitting, and photon combining.

*4.1 Beam-splitting:*

The nanoantenna-waveguide metastructure is readily extended to include the beam-splitting component by creating Y-shaped junctions at the end of the waveguide of the DBB array as shown in Fig. 5(a). The designed splitter structure shown in Figure 5(a) constitutes a splitting angle of $60^0$. To preserve efficient coupling, a cylindrical DBB is placed at the junction with a diameter of 230nm that also possesses the magnetic dipole mode at 980nm to enable symmetrical and efficient coupling between the arrays forming the nanoantenna-waveguide and the DBB array forming the branches. Finite element method based calculation of the E-field distribution of the resultant collective Mie mode at the beam-splitter region is shown in Fig. 5(b). In Fig. 5(c) we show the cross sectional Poynting vector distribution at the main waveguide part before the splitting as well as on the two branches after the splitting. As seen from the E-field distribution in Fig. 5(b) and Poynting vector distribution cross section at the end of two Y-shaped branches in Fig.5(c), the photons propagating down the waveguide get equally split in to two Y-shaped branches indicating the achieved 50/50 beam splitting. However, there is an overall 50% scattering loss at the junction (the position of the cylindrical DBB). Further optimization of the splitting efficiency can be realized by tuning the shape of the DBB at the junction. Efforts on this front is undergoing.

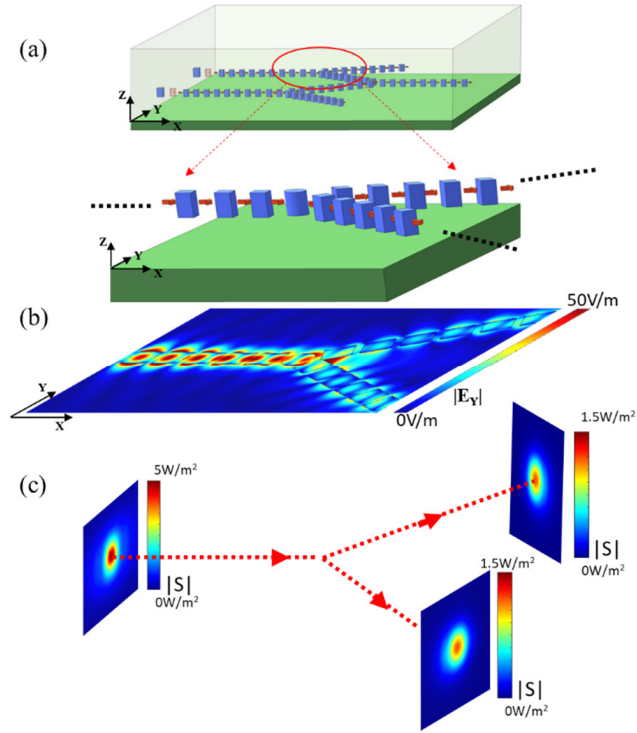

Figure 5. (a) Schematic showing the beam-splitting Y-junction of the optical circuit. The junction comprises of a DBB of cylindrical shape of diameter 230nm. The rectangular DBB size and pitch are identical to the nanoantenna-waveguide structure discussed before. (b) E-field distribution of the collective Mie mode in the spatial region of the beam-splitter that participates in the equal splitting of the photons into the two branches. (c) Cross sectional Poynting vector distribution of the input waveguide section and the two output branches of the beamsplitter.

### *4.2 Beam-Combining and Photon Interference:*

The beam-splitting component in the device architecture shown in Figure 5 can be further seamlessly extended to include the beam-combining function and thus enable interference between two photons emitted from the two distinct SPSs (shown in red and purple arrows in Fig. 6(a)). The beam-combining function is realized also via creation of Y-shaped junctions (same as used for beam splitting) to merge the photons in two different branches into the same common branch.

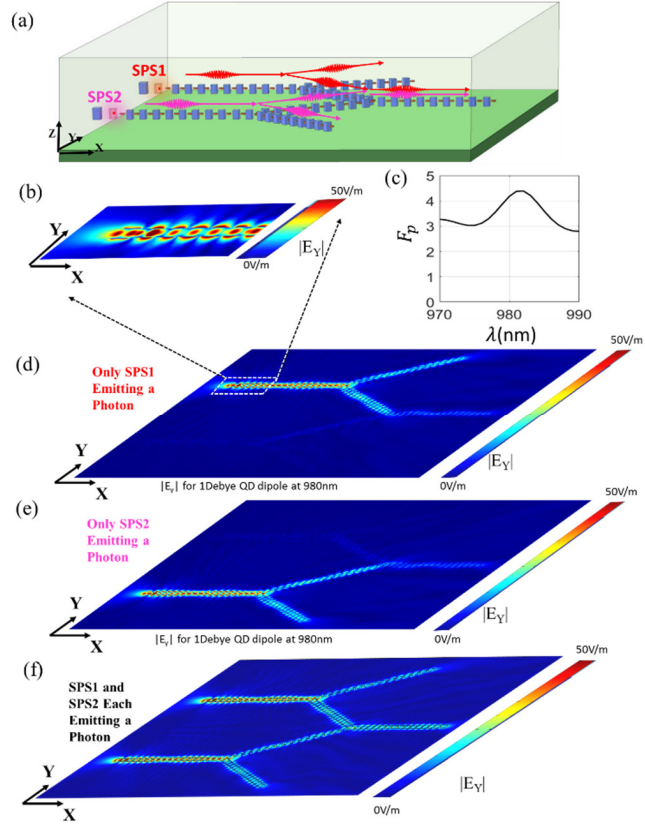

Figure 6. (a) Nanoantenna-waveguide-beamsplitting-combining circuit based on rectangular DBBs. (b) Distribution of $E_y$ around the SPS1 when only SPS1-transition dipole is emitting. (c) Purcell enhancement spectrum when only SPS1 is emitting. (d) The $E_y$ distribution of the collective Mie resonance of the whole unit when the only SPS1 is emitting: showing the splitting of the photon at the Y-junction. (e) The $E_y$ distribution of the collective Mie resonance of the unit when the both the SPSs are emitting—showing recombining of the photons in the common branch.

In such a SPS-nanoantenna-waveguide-beam splitting-beam combining unit (the primitive LMU for creating circuits) the E-field distribution, as shown in Figure 6(b), is enhanced at the location of the SPS (at a distance $\Delta L = 50 nm$ from the center of the DBB that embeds the SPS), resulting in Purcell enhancement around 4.5 at 980nm as shown in Figure 6(c). Although this enhancement is less than the value ~7 for the case of nanoantenna-waveguide segment alone shown in Fig. 3(d), the same collective Mie mode of the whole unit also propagates and splits the photons on-chip and recombines them into the common branch. This is represented in Fig. 6(d) that shows the E-field distribution of the complete network when only SPS1-transition dipole has emitted into the LMU collective mode. Similarly, Figure 6(e) shows the field distribution when only SPS2-transition dipole has emitted into the collective Mie mode. When both SPS1 and SPS2 emit simultaneously, interference of the photons takes place in the common branch. Classically the electric field distribution corresponding to such a two-photon state can be represented by superposition of the electric fields produced by the individual SPSs. Such combined electric field distribution corresponding to the in-phase interference of the two photons is shown in Fig. 6(f), illustrating constructive interference of the two photons in the common branch of the beam-combining unit. As both the photons are merged into the same collective Mie resonance of the now even larger metastrucutres, the identity of their origin is lost. Thus, measurement of a single photon out of the two-photon state will result in path-entanglement between the two SPSs [38]. Realization of such entangled states in the SPS-DBB optical circuits built out of coupled primitives comprising the SPS and a five-function-providing LMU constitutes the first step towards realizing quantum optical circuits.

## 5. Summary:

In this work we have presented an approach for design and realization of on-chip fabricable quantum optical circuits comprising regular array of SPSs with each SPS embedded in multifunctional LMU working on a single collective Mie resonance of co-designed subwavelength scale DBB based metastructure that, as a function of spatial location of the photon, provides simultaneously the necessary five photon manipulating functions of enhancement of photon emission rate, introducing directionality of emission (together, a local nanoantenna), state-preserving wave-guiding, beam-splitting, and beam-combining, enables realizing the higher level interconnected network to realize controlled interference between photons from different known sources. Based on this conceptual picture introduced in [29], we have presented here finite element method-based simulations of the optical response of all dielectric metastructures representing the above noted individual primitive and optical circuit exploiting the collective magnetic dipole Mie resonance of rectangular shaped DBBs that can be readily fabricated using nanolithography. Furthermore, we discussed a possible scheme to realistically couple efficiently an on-chip SPS to the collective Mie mode by embedding the SPS into one of the DBBs and enhancing its coupling strength to the Mie mode of that single DBB. Specifically, we have shown that, by aligning the position of the SPS with the E-field maximum of the magnetic dipole mode, the coupling strength between the SPS and a single DBB can be enhanced to lead to a Purcell enhancement ~5.5. With this scheme, we presented design and simulation results of a nanoantenna-waveguide unit that exploits this enhanced coupling of the SPS with the single DBB mode to while also providing effective coupling with the single collective mode of the array—resulting in an SPS radiative rate enhancement by ~7, directing of emitted photons to the collective mode with a collection efficiency of ~60% and lossless propagation of the photons on-chip. Furthering the move towards a complete circuit, we next showed design and simulation results of a scheme that includes a Y-junction beam splitting structure that enables 50/50 splitting of the photons in the waveguide and then recombining photons from two adjacent branches in a connected network that can allow information exchange between two distinct SPSs mediated by the collective Mie resonance of the network. Further optimization of the shape of the DBB at the splitting and recombining Y-junctions is required to achieve near-unity splitting efficiency.

The nanoantenna-waveguide-beamsplitter-beamcombiner architecture demonstrated here was chosen as a simplest embodiment of all of the five needed light manipulating functions for the SPS-DBB based quantum optical circuit design. Our approach can be readily extended to more complex architectures, e.g, Mach-Zehnder-type and Hong-Ou-Mandel-type on-chip interferometer needed for QIP, whose basic building structures are waveguides, beamsplitters and beamcombiners. The unique feature of using only a single collective Mie mode of the entire circuit where the different spatial region of the circuit provides different needed light manipulating functions eliminates the mode matching problem and hence enables ready construction of hierarchically complex SPS-DBB on-chip networks for quantum information processing.

## 6. Funding

This work was supported by Army Research Office (ARO), Grant# W911NF-15-1-0298.